# Role of Matrix Factorization Model in Collaborative Filtering Algorithm: A Survey

Dheeraj kumar Bokde[1], Sheetal Girase[2], Debajyoti Mukhopadhyay[3]
[1]Research Scholar, [2]Assitant Professor, [3]Professor and Head of Dept. of IT
Department of Information Technology, Maharashtra Institute of Technology, Pune 411038, India
dheerajbokde21@gmail.com, girase.sheetal@gmail.com, debajyoti.mukhopadhyay@gmail.com

**A B S T R A C T**

Recommendation Systems apply Information Retrieval techniques to select the online information relevant to a given user. Collaborative Filtering (CF) is currently most widely used approach to build Recommendation System. CF techniques uses the user' behavior in form of user-item ratings as their information source for prediction. There are major challenges like sparsity of rating matrix and growing nature of data which is faced by CF algorithms. These challenges are been well taken care by Matrix Factorization (MF). In this paper we attempt to present an overview on the role of different MF model to address the challenges of CF algorithms, which can be served as a roadmap for research in this area.

## I.     INTRODUCTION

Collaborative Filtering is the most popular approach to build Recommendation System and has been successfully employed in many applications. The CF recommender system works by collecting user feedback in the form of ratings for items in a given domain [1]. The most common types of CF systems is user-based and item-based approaches. The key advantage of CF recommender system is that it does not rely on the machine analyzable contents and therefore it is capable of accurate recommendations. In CF, user who had similar choices in the past, will have similar choices in the future as well.

Early collaborative filtering algorithms for recommendation systems utilize the association inferences, which have a very high time complexity and a very poor scalability. Recent methods that use matrix operations are more scalable and efficient. The implementations and algorithms of collaborative filtering for the applications of recommendation systems face several challenges. First is the size of processed datasets. The second one comes from the sparseness of rating matrix, which means for each user only a relatively small number of items are rated. So, these challenges are been well taken care by Matrix Factorization [1][2].

The Matrix Factorization (MF) plays an important role in the Collaborative Filtering recommender system. MF have recently received greater exposure, mainly as an unsupervised learning method for latent variable decomposition and dimensionality reduction [2][5]. Prediction of ratings and Recommendations can be obtained by a wide range of algorithms, while Neighborhood-based Collaborative Filtering methods are simple and intuitive. The Matrix Factorization techniques are usually more effective because they allow use to discover the latent features underlying the interactions between users and items. Matrix Factorization is simply a mathematical tool for playing around with matrices, and is therefore applicable in many domains where one would like to find out something hidden under the data. SVD and PCA are well known Matrix Factorization models for identifying latent factors in the field of Information Retrieval to deal with Collaborative Filtering challenges [2][5][6].

The rest of this paper is organized as follows. In Section II, the background of Collaborative Filtering (CF), CF techniques, Matrix Factorization (MF) is presented. In Section III, MF models are discussed. In Section IV, Role of MF model like SVD, PCA and PMF in CF algorithms is presented. The conclusion is given in the Section V.

## II. BACKGROUND

This section presents the overview of Collaborative Filtering (CF) and Matrix Factorization (MF) with their methodologies. Generally collaborative filtering techniques are classified as: Memory-Based, Model-Based and Hybrid approach.

### A. Collaborative Filtering (CF)

The term Collaborative Filtering (CF) was first coined by David Goldberg et al. [3] in 1992 to describe an email filtering system called "Tapestry". Tapestry was an electronic messaging system that allows users' to rate messages "good" or "bad" or associate text annotations with those messages. In a recommendation application, CF system tries to find other like-minded users and then recommends the items that are most liked by them based on opinions of other users.

The explosive growth of Internet usage has made the issue of information search and selection of items a very tedious task for the users, demands more efficient and scalable algorithms and implementations. For large and complex data, CF methods frequently give better performance and accuracy than Content-Based technique of Recommendation System [1][4]. Earlier Collaborative Filtering (CF) algorithms for recommendation systems used to utilize the association inferences, which have a very high time complexity and a very poor scalability. Recent methods make use of matrix operations which are more scalable and efficient.

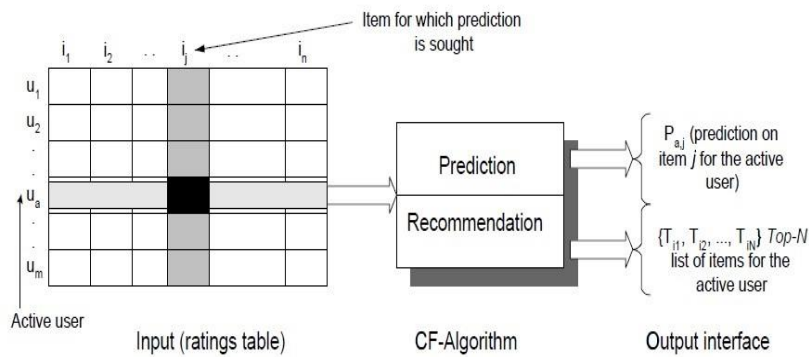

**Figure 1. The Collaborative Filtering Process**

The task of CF algorithm is to find an item likeliness that can be well described by schematic diagram of collaborative filtering process shown in Figure 1 [4]:

- Predict a numerical value *Paj* expressing the predicted score of an item *'j'* for the user *'a'*. The predicted value is within the same scale that is used by all users for rating
- Recommend a list of *Top-N* items that the active user will like the most

*Collaborative Filtering (CF) Techniques*

The Collaborative Filtering (CF) technique can be classified into following categories as follows:

### 1. Memory-Based Collaborative Filtering

The Memory-based method uses user to user and item to item correlations based on rating behavior to predict ratings and recommend items for the users in future also called as Neighborhood-Based CF. This mechanism uses users' rating data to compute similarity between users and/or items is used for making recommendations. Memory-Based CF mechanism is used in many commercial systems as it is easy to implement and is effective [1][4].

### 2. Model-Based Collaborative Filtering

Model-Based Collaborative Filtering algorithm uses RS information to create a model that generates the recommendations. Unlike Memory-Based CF, Model-based CF does not use the whole dataset to compute predictions for real data. There are various model-based CF algorithms including Bayesian Networks, Clustering Models, and Latent Semantic Models such as Singular Value Decomposition (SVD), Principal Component Analysis (PCA) and Probabilistic Matrix Factorization for dimensionality reduction of rating matrix. The goal of this approach are to uncover latent factors that explains observed ratings [1][4].

### 3. Hybrid Collaborative Filtering

To overcome the drawbacks of Memory-Based and Model-Based CF like sparsity and grey sheep are handled by these algorithm. Hybrid Collaborative Filtering algorithms are the combination of Memory-Based and Model-Based Collaborative Filtering approaches. It improves the prediction performance of the CF algorithms [1].

### B. Matrix Factorization (MF)

Most of the MF models are based on the latent factor model [2]. Matrix Factorization approach is found to be most accurate approach to reduce the problem from high levels of sparsity in RS database, certain studies have used dimensionality reduction techniques. In the model-based technique Latent Semantic Index (LSI) and the dimensionality reduction method Singular Value Decomposition (SVD) are typically combined [2][4]. SVD and PCA are well-established technique for identifying latent factors in the field of Information Retrieval to deal with CF challenges. These methods have become popular recently by combining good scalability with predictive accuracy. They offers much flexibility for modeling various real-life applications.

Firstly, we have a set of U users, and a set of I items. Let R be the matrix of size |U| x |I| that contains all the ratings that the users have assigned to the items. Now the latent features would be discovered. Our task then, is to find two matrices, P (|U|x K) and Q (|I|x K) such that their product approximately equals to R is given by:

$$R \approx P \times Q^T = \hat{R}$$

In this way, the Matrix factorization models map both users and items to a joint latent factor space of dimensionality *f*, user-item interactions are modeled as inner products in that space [2]. Accordingly, each item *i* is associated with a vector $q_i \in R^f$, and each user *u* is associated with a vector $p_u \in R^f$. For a given item *i*, the elements of $q_i$ measure the extent to which the item possesses those factors positive or negative. The resulting dot product $q_i^T p_u$ captures the interaction between user *u* and item *i*, the users' overall interest in the item characteristics. This approximates user *u*'s rating of item *i* which is denoted by $r_{ui}$ leading to the estimate: [2]:

$$\hat{r}_{ui} = q_i^T p_u$$

To learn the factor vectors ($p_u$ and $q_i$), the system minimizes the regularized squared error on the set of known ratings as [2]:

$$\min_{q^*, p^*} \sum_{(u,i) \in K} (r_{ui} - q_i^T p_u)^2 + \lambda(\|q_i\|^2 + \|p_u\|^2)$$

Here, K is the set of the (*u, i*) pairs for which $r_{ui}$ is known the training set. The constant λ controls the extent of regularization and is usually determined by cross-validation.

### III. MATRIX FACTORIZATION (MF) MODELS

There are various matrix factorization models, some commonly used are:

#### A. Singular Value Decomposition (SVD)

The Singular Value Decomposition (SVD) is the powerful technique of dimensionality reduction. The key issue in an SVD decomposition is to find a lower dimensional feature space.

SVD of an $m \times n$ matrix **A** is of the form [1][5]:

$$SVD(A) = U\Sigma V^T$$

Where,
 **U** and **V** are $m \times m$ and $n \times n$ orthogonal matrices respectively
 **Σ** is the $m \times n$ singular orthogonal matrix with non-negative elements

An $m \times m$ matrix **U** is called orthogonal if $U^T U$ equals to an $m \times m$ identity matrix. The diagonal elements in **Σ** ($\sigma_1, \sigma_2, \sigma_3, \ldots \sigma_n$) are called the singular values of matrix **A**. Usually, the singular values are placed in the descending order in Σ. The column vectors of **U** and **V** are called the left singular vectors and the right singular vectors respectively [5].

#### B. Principal Component Analysis (PCA)

The Principal Component Analysis (PCA) is also the powerful technique of dimensionality reduction and is a particular realization of the Matrix Factorization (MF) approach [1][6]. PCA is a statistical procedure that uses an orthogonal transformation to convert a set of observations of possibly correlated variables into a set of values of linearly uncorrelated variables called principal components. The number of original variable is greater than or equal to principal components. This transformation is defined in such a way that a linear projection of high dimensional data into a lower dimensional subspace such as the variance retained is maximized and the least square reconstruction error is minimized. The principal components are orthogonal because they are the eigenvectors of the covariance matrix. PCA is sensitive to the relative scaling of the original variables.

PCA allows to obtain an ordered list of components that account for the largest amount of the variance from the data in terms of least square errors. The amount of variance captured by the first component is larger than the amount of variance on the second component and so on. We can reduce the dimensionality of the data by neglecting those components.

#### C. Probabilistic Matrix Factorization (PMF)

The Probabilistic Matrix Factorization (PMF) is a probabilistic linear model with Gaussian observation noise [7]. The user preference matrix is represented as the product of tow lower-rank user and item matrices in Probabilistic Matrix Factorization (PMF). Suppose we have $N$ users and $M$ movies. Let $R_{ij}$ be the rating value of user $i$ for movie $j$, $U_i$ and $V_j$ represent D-dimensional user-specific and movie-specific latent feature vectors respectively.

Then the conditional distribution over the observed ratings $R \in \mathbb{R}^{N \times M}$ and the prior distributions over $U \in \mathbb{R}^{D \times N}$ and $V \in \mathbb{R}^{D \times M}$ are given by in [7]:

$$p(R|U,V,\sigma^2) = \prod_{i=1}^{N}\prod_{j=1}^{M}\left[\mathcal{N}(R_{ij}|U_i^T V_j, \sigma^2)\right]^{I_{ij}}$$

$$p(U|\sigma_U^2) = \prod_{i=1}^{N}\mathcal{N}(U_i|0,\sigma_U^2 I)$$

$$p(V|\sigma_V^2) = \prod_{j=1}^{M}\mathcal{N}(V_j|0,\sigma_V^2 I)$$

Where,
 $\mathcal{N}(x|\mu,\sigma^2)$ denotes the Gaussian distribution with mean **μ** and variance $\sigma^2$
 $I_{ij}$ is the indicator variable that is equal to **1** if user *i* rated movie *j* and equal to **0** otherwise

## IV. ROLE OF MATRIX FACTORIZATION IN COLLABORATIVE FILTERING ALGORITHM

Collaborative Filtering is a most promising research field in the area of Information Retrieval, so many researchers have contributed to this area. Many CF researchers have recognized the problem of large dataset and sparseness (i.e., many values in the ratings matrix are null since all users do not rate all items), which is been well taken care by Matrix Factorization. Computing distances between users is complicated by the fact that the number of items users have rated in common is not constant. It is important to study the role of Matrix Factorization models like SVD, PCA and PMF with Collaborative Filtering (CF) algorithms.

Looking at the contribution of other researchers who have worked in this area have motivated us to work on the role of Matrix Factorization model in Collaborative Filtering algorithm. An overview of research work done by other researchers is presented as a survey in this section below.

Badrul Sarwar et al. [4] explored Item-Based Collaborative Filtering technique to produce high quality recommendations. They first analyzed the user-item rating matrix to identify the relationship between different items, and then used these relationship to directly compute recommendations for the users by using Item-based technique. They applied Matrix Factorization model SVD to reduce the dimensionality of a ratings matrix. Using the MovieLens dataset, they selected 943 users to form a (943 × 1682) matrix each user on average rates 5% of the 1682 movies i.e., 95.4% sparse. They first fill missing values using user and movie rating averages, and then apply SVD. For this large dataset Item-Based technique provide optimal accuracy with significantly faster and high quality online recommendations than user-user (k-Nearest-Neighbor) method.

Goldberg et al. [8] proposed an approach to use Principal Component Analysis (PCA) in the context of an online Joke Recommendation System. Their system, known as Eigentaste [8]. In Eigentaste they addressed sparseness using universal queries, which insure that all users rate a common set of k-items. So, resulting sub-matrix will be dense and directly compute the square symmetric correlation matrix and then did linear projection using Principle Component Analysis (PCA), a closely-related factor analysis technique first described by Pearson in 1901. Like SVD, PCA reduces dimensionality of matrix by optimally projecting highly correlated data along a smaller number of orthogonal dimensional subspace such as the variance retained is maximized and the least square reconstruction error is minimized.

Royi Ronen et al. [9] proposed a project Sage, Microsoft's all-purpose recommender system designed and developed as an ultra-high scale cloud service. The main focus of project Sage is on both state of the art research and high scale robust implementation. A novel Probabilistic Matrix Factorization (PMF) model was presented by Royi et al. for implicit one-class data as new evaluation framework. Their service Sage is deployed on the Microsoft Azure cloud which provides easy to use interface to integrate a recommendation service into any website. Recommender Systems based on Matrix Factorization (MF) models have repeatedly demonstrated better accuracy than other methods, such as Nearest-Neighbor models and restricted Boltzmann machines. The dashboard allows users to choose any subset of items and generate high quality recommendations as well as explore item-to-item relation.

## V. CONCLUSION

Collaborative Filtering (CF) algorithms are most commonly used in Recommendation System (RS). CF algorithms now a day face the problem with large dataset and sparseness in rating matrix. In this paper we have studied role of various Matrix Factorization (MF) model to deal with the Collaborative Filtering (CF) challenges.

From this study we can say that, SVD is able to handle large dataset, sparseness of rating matrix and scalability problem of CF algorithm efficiently. PCA is able to finds a linear projection of high dimensional data into a lower dimensional subspace such as the variance retained is maximized and the least square reconstruction error is minimized. All the techniques which have been researched until now are trying to increase the accuracy and prediction performance of CF algorithm by dimensionality reduction of rating matrix using MF model.